\documentclass[conference,a4paper]{IEEEtran}
\usepackage{xcolor}
\usepackage{balance}

\usepackage{cite}
\usepackage{amsmath,amssymb,amsfonts}
\usepackage{graphicx}
\usepackage{textcomp}
\usepackage{acronym}
\usepackage{xcolor}
\usepackage{tikz} % for tikz
\usepackage[utf8]{inputenc}
\usepackage{pgfplots} 
\usepackage{pgfgantt}
\usepackage{pdflscape}
\usepackage{changes}
\usepackage{comment}
\usepackage{subfigure}

\usepackage{mathtools,algpseudocode,algorithm,MnSymbol}
\acrodef{MMSE}{Minimum Mean Squared Error}

\acrodef{MSE}{mean square error}

\acrodef{PSD}{power spectral density}

\acrodef{RMSE}{root mean squared error}
\acrodef{SLR}{statistical linear regression}

\acrodef{IPLF}{iterated posterior linearization filter}

\acrodef{ue}[UE]{user equipment}
\acrodef{bs}[BS]{base station}
\acrodef{va}[VA]{virtual anchor}
\acrodef{sp}[SP]{scattering  point}
\acrodef{fov}[FoV]{field-of-view}   
\acrodef{los}[LOS]{line-of-sight}
\acrodef{nlos}[NLOS]{non-line-of-sight}
\acrodef{PMBM}[PMBM]{Poisson  multi-Bernoulli  mixture}
\acrodef{PMB(M)}[PMB(M)]{Poisson  multi-Bernoulli  (mixture)}
\acrodef{PMB}[PMB]{Poisson  multi-Bernoulli}
\acrodef{rfs}[RFS]{random finite set}
\acrodef{PPP}[PPP]{Poisson point process}
\acrodef{MBM}[MBM]{multi-Bernoulli  mixture}
\acrodef{ekf}[EKF]{extended Kalman filter}
\acrodef{PDF}[PDF]{probability density function}

\acrodef{ckf}[CKF]{cubature Kalman filter}
\acrodef{rbp}[RBP]{Rao-Blackwellized particle}
\acrodef{gospa}[GOSPA]{generalized optimal subpattern assignment}
\acrodef{slam}[SLAM]{simultaneous localization and mapping}

\acrodef{TOA}[TOA]{time of arrival}
\acrodef{AOA}[AOA]{angles of arrival}
\acrodef{AOD}[AOD]{angles of departure}
 
\usepackage{pgfplots}
  \pgfplotsset{compat=newest}
  \usetikzlibrary{plotmarks}
  \usetikzlibrary{arrows.meta}
  \usepgfplotslibrary{patchplots}
  \usepackage{grffile}
  \usepackage{amsmath}

\pgfplotsset{compat=newest} 
\pgfplotsset{plot coordinates/math parser=false} % end of tikz

\begin{document}

\bibliographystyle{IEEEtran}
\bstctlcite{IEEEexample:BSTcontrol}
%
% paper title
% Titles are generally capitalized except for words such as a, an, and, as,
% at, but, by, for, in, nor, of, on, or, the, to and up, which are usually
% not capitalized unless they are the first or last word of the title.
% Linebreaks \\ can be used within to get better formatting as desired.
% Do not put math or special symbols in the title.
%\title{An End-to-end 5G SLAM Framework with a Low-complexity Channel Estimator}
\title{Iterated Posterior Linearization PMB Filter\\ for 5G SLAM}

% author names and affiliations
% use a multiple column layout for up to three different
% affiliations
\author{\IEEEauthorblockN{
Yu Ge\IEEEauthorrefmark{1},   % 1st author, 1st affiliations
Yibo Wu\IEEEauthorrefmark{1},
Fan Jiang\IEEEauthorrefmark{1},
Ossi Kaltiokallio\IEEEauthorrefmark{2},\\
Jukka Talvitie\IEEEauthorrefmark{2},
Mikko Valkama\IEEEauthorrefmark{2},
Lennart Svensson\IEEEauthorrefmark{1},      % 4th author, 4th affiliations
Henk Wymeersch\IEEEauthorrefmark{1}      % 4th author, 4th affiliations
}                                     % ...
%\\
\IEEEauthorblockA{\IEEEauthorrefmark{1}% 1st affiliations
Department of Electrical Engineering, Chalmers University of Technology, Gothenburg, Sweden,\\ 
\IEEEauthorblockA{\IEEEauthorrefmark{2}% 2nd affiliations
Unit of Electrical Engineering, Tampere University, Tampere, Finland,\\ }
\{yuge,~yibo,~fan.jiang,~lennart.svensson,~henkw\}@chalmers.se, \{ossi.kaltiokallio,jukka.talvitie,mikko.valkama\}@tuni.fi}
}

% conference papers do not typically use \thanks and this command
% is locked out in conference mode. If really needed, such as for
% the acknowledgment of grants, issue a \IEEEoverridecommandlockouts
% after \documentclass

% use for special paper notices
%\IEEEspecialpapernotice{(Invited Paper)}

% make the title area
\maketitle

% As a general rule, do not put math, special symbols or citations
% in the abstract
\begin{abstract}
5G millimeter wave (mmWave) signals have inherent geometric connections to the propagation channel and the propagation environment. Thus, they can be used to jointly localize the receiver and map the propagation environment, which is termed as \ac{slam}. One of the most important tasks in the 5G \ac{slam} is to deal with the nonlinearity of the measurement model. To solve this problem, existing 5G SLAM approaches rely on sigma-point or extended Kalman filters, linearizing the measurement function with respect to the \emph{prior} \ac{PDF}. In this paper, we study the linearization of the  measurement function with respect to the \emph{posterior} \ac{PDF}, and implement the iterated posterior linearization filter into the Poisson multi-Bernoulli  SLAM filter. Simulation results demonstrate the accuracy and precision improvements of the resulting SLAM filter. 
\end{abstract}

\vskip0.5\baselineskip

% For peer review papers, you can put extra information on the cover
% page as needed:
% \ifCLASSOPTIONpeerreview
% \begin{center} \bfseries EDICS Category: 3-BBND \end{center}
% \fi
%
% For peerreview papers, this IEEEtran command inserts a page break and
% creates the second title. It will be ignored for other modes.
% \IEEEpeerreviewmaketitle

\section{Introduction}
5G mmWave signals provide unique opportunities for \acf{slam}, due to their inherent geometric connection to the propagation environment \cite{nurmi2017multi}. Signals from the base station (BS) reach the \ac{ue} via multiple propagation paths. Each path is determined by the propagation environment and the locations of the BS and the \ac{ue}. State-of-the-art channel estimators can provide accurate estimates for those paths by using received signals, in terms of groups of channel gain, \ac{TOA}, \ac{AOA}, and \ac{AOD}, which contain information needed for \ac{slam} \cite{witrisal2016high,wymeersch20175g}. 

Much work has been done in SLAM using 5G signals, also called 5G SLAM, including geometry-based methods \cite{wen20195g, yassin2018mosaic}, which cannot provide uncertainty information;  message passing methods \cite{mendrzik2018joint,Erik_BPSLAM_TWC2019,Rico_BPSLAM_JSTSP2019}, which provide uncertainty information, but are inherently sub-optimal; more powerful algorithms using \ac{rfs} theory  \cite{kim20205g,kim2020low,ge2020exploiting,ge20205GSLAM,ge2021computationally}. The latter class of algorithms can handle the data association (DA) problem between measurements and landmarks \cite{bar1990tracking} and have certain optimality guarantees. In particular, probability hypothesis density (PHD) filters are used in \cite{kim20205g,kim2020low}, \ac{PMBM} filters are used in \cite{ge2020exploiting,ge20205GSLAM,ge2021computationally}, and the  low-computational version \ac{PMB} filters are also applied in \cite{ge2021computationally}. Because PMB(M) filters enumerate all possible DAs explicitly, they provide more accurate results than PHD filters in 5G SLAM, as shown in \cite{ge2021computationally}.
\begin{figure}
\centering
    \includegraphics[width=0.8\linewidth]{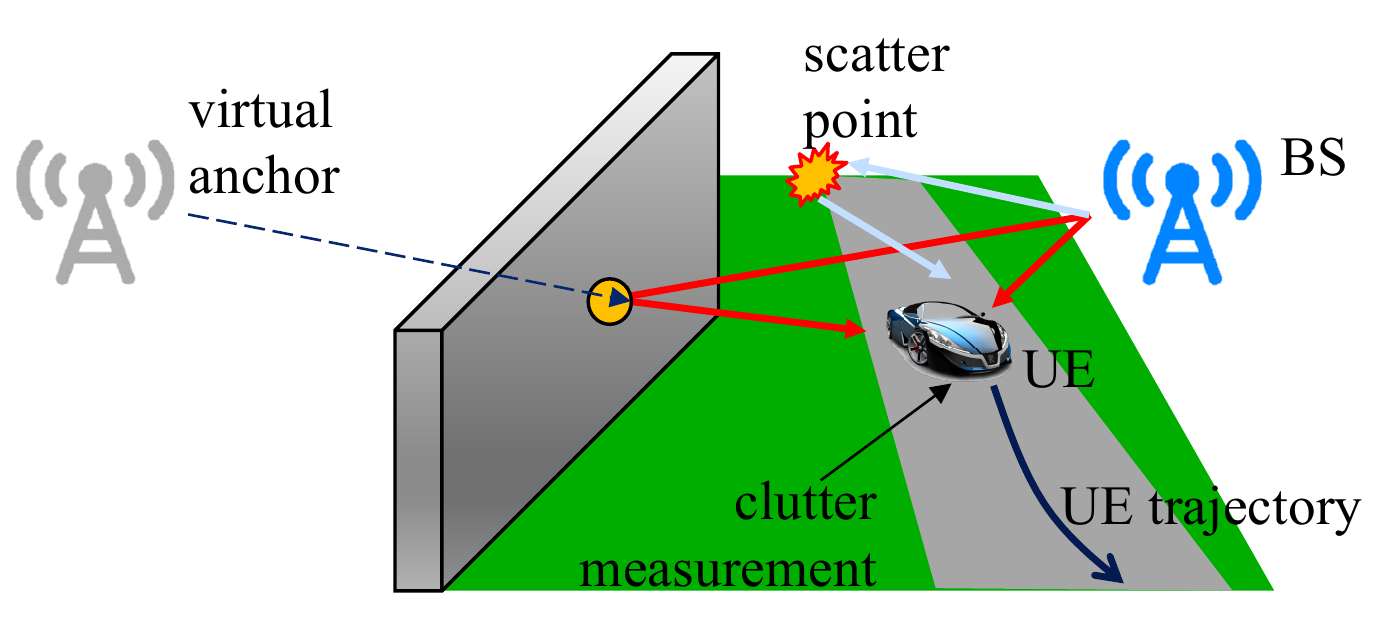}
\caption{5G downlink scenario with the environment of a \ac{bs}, a \ac{ue}, a reflecting surface, and a scatter point, where the \ac{ue} can not only track its own state, but also construct the map of the surrounding landmarks, by using channel parameters estimated from received downlink signals.}
\label{fig:overview}
%\vspace{-4mm}
\end{figure}

Common to all 5G SLAM studies, it is the requirement to account for the nonlinearity of the measurement function, which in the case of 5G mmWave signals relate \ac{TOA}, \ac{AOA}, and \ac{AOD}, to positions and orientations of the \ac{ue} and landmarks. To this end, \cite{ge2020exploiting,ge20205GSLAM} follow a Rao-Blackwellized approach, and utilize the particle filter (for the \ac{ue} state) \cite[Ch. 7.1]{sarkka2013bayesian}, and the \ac{ckf} (for the landmark states, conditioned on the \ac{ue} state) which uses sigma-points drawn from the prior \acf{PDF}  \cite{arasaratnam2009cubature} to propagate through the measurement nonlinearity \cite[Ch. 5.2]{sarkka2013bayesian}. In \cite{ge2021computationally},  the \ac{ekf} is implemented into the PMB(M) \ac{slam} filter, and the approximation of the nonlinearity is formed by utilizing the first-order  Taylor  series \cite[Ch. 5.2]{sarkka2013bayesian}, which is equivalent to 
a linearization at the prior mean. However, these methods either have high computational burden, or perform poorly with nonlinear measurement functions, if the measurement noise is low enough \cite{morelande2013analysis}. A more accurate and efficient linearization method is proposed in \cite{garcia2015posterior,garcia2021gaussian}, which linearizes the measurement function with respect to the posterior \ac{PDF}, rather than the prior \ac{PDF}. To our best knowledge, the evaluation of such an approach in 5G SLAM has not yet been conducted. 

In this paper, we show that the linearization of the measurement function can be done with respect to the posterior joint \ac{PDF} of the \ac{ue} state and the landmark, and extend our previous work in \cite{ge2021computationally} to implement the posterior linearization into the PMB-based 5G SLAM filter. The main contributions of this paper are summarized as follows: (i) we develop the \ac{IPLF} integrated to the 5G PMB \ac{slam} filter; (ii) we show that the proposed IPL-PMB SLAM filter can improve the mapping and positioning accuracy and precision, while guaranteeing near real-time operation, although a minor computational cost needs be paid.

\subsubsection*{Notations}
Scalars (e.g., $x$) are denoted in italic, vectors (e.g., $\boldsymbol{x}$) in bold lower-case letters, matrices (e.g., $\boldsymbol{X}$) in bold capital letters, sets  (e.g., $\mathcal{X}$) in calligraphic. Transpose is denoted by $(\cdot)^{\mathsf{T}}$, the union of mutually disjoint sets is denoted by $\uplus$, a Gaussian density with mean $\boldsymbol{u}$ and covariance $\boldsymbol{\Sigma}$, evaluated in value $\boldsymbol{x}$, is denoted by $\mathcal{N}(\boldsymbol{x};\boldsymbol{u},\boldsymbol{\Sigma})$, and $d_{\boldsymbol{x}}=\text{dim}(\boldsymbol{x})$.

%\vspace{-1mm}
\section{System model}
In this section, the \ac{ue} model, the environment model, and the measurement model for a 5G downlink localization scenario, as shown in Fig. \ref{fig:overview}, are introduced.

%\vspace{-1mm}
\subsection{User Model}
A single-user scenario is considered, thus the cooperation among \acp{ue} is out of scope of this paper. We denote the dynamic state of the \ac{ue} at time step $k$ as $\boldsymbol{s}_{k}$, which at least contains the \ac{ue} position $\boldsymbol{x}_{\mathrm{UE},k}=[x_{k},y_{k},z_{k}]^{\mathsf{T}}$, the heading $\varpi_{k}$ and the clock bias $B_{k}$. If the process noise is zero-mean Gaussian, the transition density of $\boldsymbol{s}_{k}$ can be expressed as 
\begin{equation}
f(\boldsymbol{s}_k | \boldsymbol{s}_{k-1}) = {\cal N}(\boldsymbol{s}_k ; \boldsymbol{v}(\boldsymbol{s}_{k-1}),\boldsymbol{Q}_{k-1}), \label{transition_density}
\end{equation}
where $\boldsymbol{v}(\cdot)$ denotes a known transition function, and $\boldsymbol{Q}_{k-1}$ denotes a known covariance matrix.

\subsection{Environment Model}
We consider an environment with three different types of landmarks, which are the \ac{bs}, reflecting surfaces, and small objects. In the environment, there is a \ac{bs} with known position, which sends downlink signals to the \ac{ue}, and a few unknown reflecting surfaces and small objects. They can reflect and scatter the downlink signals to the \ac{ue}, respectively,  and are modeled as \acp{va} and \acp{sp} (see Fig.\,\ref{fig:overview}). We model the landmark state  as $\boldsymbol{x} = [\boldsymbol{x}^{\textsf{T}}_{\text{LM}},m]^{\textsf{T}}$,  where $\boldsymbol{x}_{\text{LM}} \in \mathbb{R}^{3}$ represents the landmark location, and $m\in\{\text{BS},\text{VA},\text{SP}\}$ represents the landmark type. Therefore, the map of the environment can be represented by a set of landmark $\mathcal{X}=\{\boldsymbol{x}^{1},\dots, \boldsymbol{x}^{I}\}$, with ${I}$ representing the total number of landmarks.

\subsection{Measurement Model} \label{Measurement model}
At time step $k$, the \ac{ue} receives downlink signals from the \ac{bs}. When considering OFDM transmissions, we can express the received signal at subcarrier $\kappa$ at time step $k$ as~\cite{heath2016overview}
\begin{align}
    &\boldsymbol{Y}_{\kappa,k} =\boldsymbol{C}_{\kappa,k} \boldsymbol{S}_{\kappa} + \boldsymbol{N}_{\kappa,k}, \label{eq:FreqObservation} 
\end{align}
where $\boldsymbol{S}_{\kappa}$ is the (possibly pre-coded) pilot signal over subcarrier $\kappa$, $\boldsymbol{Y}_{\kappa,k}$ is the received signal over subcarrier $\kappa$, $\boldsymbol{N}_{\kappa,k}$ is white Gaussian noise, and $\boldsymbol{C}_{\kappa,k}$ is the channel frequency response. As the transmitted signals can reach the \ac{ue} directly, which is the \ac{los} path, and/or reflected by reflecting surfaces or scattered by small objects, which are \ac{nlos} paths, $\boldsymbol{C}_{\kappa,k}$ can be denoted as
\begin{align}
    &\boldsymbol{C}_{\kappa,k} = \boldsymbol{W}_{k}^{\mathsf{H}}\sum _{i=0}^{I_{k}-1}g_{k}^{i}\boldsymbol{a}_{\text{R}}(\boldsymbol{\theta}_{k}^{i})\boldsymbol{a}_{\text{T}}^{\mathsf{H}}(\boldsymbol{\phi}_{k}^{i})e^{-\jmath 2\pi \kappa \Delta f \tau_{k}^{i}}, \label{eq:FreqObservation2} 
\end{align}
where $\boldsymbol{W}_{k}$ represents a combining matrix, $\boldsymbol{a}_{\text{R}}(\cdot)$ and $\boldsymbol{a}_{\text{T}}(\cdot)$ denote the steering vectors of the receiver and transmitter antenna arrays, respectively, and $\Delta f$ denotes the subcarrier spacing. Moreover, $I_{k}$ is the number of all visible landmarks, and we assume that there is only one path from each landmark. The \ac{los} path corresponds to $i=0$, and the \ac{nlos} paths to $i>0$. Each path $i$ can be described by a complex gain $g_{k}^{i}$, a \ac{TOA} $\tau_{k}^{i}$, an \ac{AOA} pair $\boldsymbol{\theta}_{k}^{i}$ in azimuth and elevation, and an \ac{AOD} pair $\boldsymbol{\phi}_{k}^{i}$ in azimuth and elevation. Those channel parameters depend on the hidden geometric relation among the \ac{bs}, \ac{ue} and landmarks, which can be found, e.g., in \cite[Appendix A]{ge20205GSLAM}. 

At the \ac{ue} side, a channel estimator, such as \cite{richter2005estimation,alkhateeb2014channel,venugopal2017channel,Gershman2010,jiang2021high}, provides estimates of angles and delays of paths from $\boldsymbol{Y}_{\kappa,k}$. However, the channel estimation is out of the scope of this paper, and the \ac{ue} directly utilizes output of the channel estimator that provides the angel and delay estimates.
%the They can be estimated by applying  channel estimation algorithm, such as  \cite{richter2005estimation,alkhateeb2014channel,venugopal2017channel,Gershman2010,jiang2021high}, on $\boldsymbol{Y}_{s,k}$.
At time step $k$, a set of measurements $\mathcal{Z}_{k}=\{\boldsymbol{z}_{k}^{1},\dots, \boldsymbol{z}_{k}^{\hat{{I}}_{k}} \}$ is provided, where usually $\hat{{I}}_{k} \neq {{I}}_{k}$, as there may be some clutter measurements and  misdetected landmarks. If the measurement noise is zero-mean Gaussian, the measurement originating from landmark $\boldsymbol{x}^{i}$ follows
\begin{align}
    f(\boldsymbol{z}_{k}^{i}|\boldsymbol{x}^{i},\boldsymbol{s}_{k})=\mathcal{N}(\boldsymbol{z}_{k}^{i};\boldsymbol{h}(\boldsymbol{x}^{i},\boldsymbol{s}_{k}),\boldsymbol{R}_k^i),\label{pos_to_channelestimation}
\end{align}
where  
$\boldsymbol{h}(\boldsymbol{x}^{i},\boldsymbol{s}_{k})=[\tau_{k}^{i},(\boldsymbol{\theta}_{k}^{i})^{\mathsf{T}},(\boldsymbol{\phi}_{k}^{i})^{\mathsf{T}}]^{\mathsf{T}}$ represents the nonlinear function that transforms the geometric information to the TOA, AOA and AOD, 
and $\boldsymbol{R}_k^i$ is the measurement covariance. 

\section{PMB(M) SLAM Filter}
%\vspace{-1mm}
In this section, we approximate the map $\mathcal{X}$ conditioned on the \ac{ue} state $\boldsymbol{s}_{k}$ as a PMB density. In other words, it is a PMB \ac{rfs}. We will now briefly introduce the  basics of the PMB(M) density and   the PMB(M) SLAM filter.

\subsection{Basics of PMB(M) Density}
The \ac{PMBM} \ac{rfs} $\mathcal{X}$ can be viewed as the union of two disjoint \acp{rfs}, $\mathcal{X}_{\mathrm{U}}$ and $\mathcal{X}_{\mathrm{D}}$, which are the set of undetected objects that have been never detected,  and the set of detected objects that have been detected at least once, respectively \cite{garcia2018poisson}. The \ac{rfs} $\mathcal{X}_{\mathrm{U}}$ is usually modeled as a \ac{PPP}, with the density following  
\begin{equation}
    f_{\mathrm{P}}(\mathcal{X}_{\mathrm{U}})=e^{-\int\lambda(\boldsymbol{x})\mathrm{d}\boldsymbol{x}}\prod_{\boldsymbol{x} \in \mathcal{X}_{\mathrm{U}} }\lambda(\boldsymbol{x}),\label{PPP}
\end{equation}
where $\lambda(\cdot)$ is the intensity function. 
The \ac{rfs} $\mathcal{X}_{\mathrm{D}}$ is usually modeled as a \ac{MBM}, with the density following
\begin{equation}
    f_{\mathrm{MBM}}(\mathcal{X}_{\mathrm{D}})\propto \sum_{j \in \mathbb{I}}w^{j}\sum_{\mathcal{X}^{1}\biguplus \dots \biguplus \mathcal{X}^{n}=\mathcal{X}_{\mathrm{D}}}\prod_{i=1}^{n}f^{j,i}_{\mathrm{B}}(\mathcal{X}^{i}),\label{MBM}
\end{equation}
where  $\mathbb{I}$ is the index set of all global hypotheses and $w^{j}\ge 0$ is the weight for $j$-th global hypothesis, satisfying $\sum_{j\in\mathbb{I}}w^{j}=1$ \cite{williams2015marginal}; $n$ is the number of potentially detected objects; $f_{\mathrm{B}}^{j,i}(\cdot)$ is the Bernoulli density of the $i$-th landmark under the $j$-th global hypothesis. Each Bernoulli follows
\begin{equation}
f^{j,i}_{\mathrm{B}}(\mathcal{X}^{i})=
\begin{cases}
1-r^{j,i} \quad& \mathcal{X}^{j}=\emptyset \\ r^{j,i}f^{j,i}(\boldsymbol{x}) \quad & \mathcal{X}^{j}=\{\boldsymbol{x}\} \\ 0 \quad & \mathrm{otherwise}
\end{cases}
\end{equation} 
where $r^{j,i} \in [0,1]$ is the existence probability, and $f^{j,i}(\cdot)$ is the state density. More details of the PPP and MBM densities can be found in \cite{williams2015marginal,garcia2018poisson,fatemi2017poisson}.
Then, the density of $\mathcal{X}$ can be computed using the convolution formula \cite[eq. (4.17)]{mahler2014advances} as
\begin{equation}
    f(\mathcal{X})=\sum_{\mathcal{X}_{\mathrm{U}}\biguplus\mathcal{X}_{\mathrm{D}}=\mathcal{X}}f_{\mathrm{P}}(\mathcal{X}_{\mathrm{U}})f_{\mathrm{MBM}}(\mathcal{X}_{\mathrm{D}}),\label{PMBM}
\end{equation}
which can also be parameterized by $\lambda(\boldsymbol{x})$ and $\{w^{j},\{r^{j,i},f^{j,i}(\boldsymbol{x})\}_{i\in \mathbb{I}^{j}}\}_{j\in \mathbb{I}}$, with $\mathbb{I}^{j}$ representing the index set of landmarks (i.e., the Bernoulli components) under the $j$-th global hypothesis.  If there is only one mixture component in the MBM, then \eqref{PMBM} reduces to a PMB.

\subsection{PMB(M) SLAM Filter}
The PMBM SLAM filter follows the prediction and update steps of the Bayesian filtering recursion with RFSs \cite{mahler2003multitarget}. In practice, instead of tracking the joint posterior  $f(\mathbf{s}_{0:k},\mathcal{X}|\mathcal{Z}_{1:k})$,   we keep track of marginal posteriors $f(\mathcal{X}|\mathcal{Z}_{1:k})$ and $f(\mathbf{s}_{k}|\mathcal{Z}_{1:k})$ to reduce complexity. %To keep track of marginal posteriors $f(\mathcal{X}|\mathcal{Z}_{1:k})$ and $f(\boldsymbol{s}_{k}|\mathcal{Z}_{1:k})$
To do this, 
the prediction of the \ac{ue} state follows the Chapman-Kolmogorov equation, given by 
\begin{align}
     f(\boldsymbol{s}_{k+1}|\mathcal{Z}_{1:k}) & = \int f(\boldsymbol{s}_{k}|\mathcal{Z}_{1:k})f(\boldsymbol{s}_{k+1}|\boldsymbol{s}_{k}) \text{d} \boldsymbol{s}_{k}.\label{predicted_prior_vehicleMarg}
\end{align} 
As all landmarks are static, there is no prediction for the map. By marginalizing out the map state in the joint posterior, the update step for the \ac{ue} state becomes
\begin{align}
   &  f(\boldsymbol{s}_{k+1}|\mathcal{Z}_{1:k+1})  = \int  f(\boldsymbol{s}_{k+1},\mathcal{X}|\mathcal{Z}_{1:k+1})\delta \mathcal{X}  \label{eq:UpMarObjMarg1}\\
    %& = \frac{1}{f(\mathcal{Z}_{k+1}|\mathcal{Z}_{1:k})}
    & \propto \int f(\mathcal{X}|\mathcal{Z}_{1:k})f(\boldsymbol{s}_{k+1}|\mathcal{Z}_{1:k}) g(\mathcal{Z}_{k+1}|\boldsymbol{s}_{k+1},\mathcal{X}) \delta \mathcal{X}, \label{eq:UpMarObjMarg3}
\end{align}
whereas by marginalizing out the \ac{ue} state, the map state follows
\begin{align}
    &f(\mathcal{X}|\mathcal{Z}_{1:k+1}) = \int f(\boldsymbol{s}_{k+1},\mathcal{X}|\mathcal{Z}_{1:k+1}) \mathrm{d}\boldsymbol{s}_{k+1}\label{eq:UpMarObjMarg2}\\
    %& = \frac{1}{f(\mathcal{Z}_{k+1}|\mathcal{Z}_{1:k})} 
    & \propto \int f(\mathcal{X}|\mathcal{Z}_{1:k})f(\boldsymbol{s}_{k+1}|\mathcal{Z}_{1:k}) \ell(\mathcal{Z}_{k+1}|\boldsymbol{s}_{k+1},\mathcal{X})
    \mathrm{d}\boldsymbol{s}_{k+1}, \label{eq:UpMarObjMarg4}
\end{align}
where $\ell(\mathcal{Z}_{k+1}|\boldsymbol{s}_{k+1},\mathcal{X})$ is the RFS likelihood function, given by  \cite[eqs.\,(5)--(6)]{garcia2018poisson}, and $\int \psi(\mathcal{X})\delta \mathcal{X}$ refers to the set integral \cite[eq.~(4)]{williams2015marginal}. In practice,  \eqref{predicted_prior_vehicleMarg}, \eqref{eq:UpMarObjMarg1}, \eqref{eq:UpMarObjMarg2} are usually translated into prediction and update steps of the PMBM parameters  $\lambda(\boldsymbol{x})$ and $\{w^{j},\{r^{j,i},f^{j,i}(\boldsymbol{x})\}_{i\in \mathbb{I}^{j}}\}_{j\in \mathbb{I}}$.
As the number of DAs increases very rapidly over time, the exact PMBM SLAM filter has high complexity.  To mitigate this, the PMB SLAM filter is often used, which approximates the PMBM density to a PMB density at the end of each time step by marginalizing over DAs. %Without loss of generality, we will consider the PMB SLAM filter from \cite{ge2021computationally}.

\subsection{EK-PMB(M) \ac{slam} Filter}
The EK-PMB(M) \ac{slam} filter from \cite{ge2021computationally} computes \eqref{eq:UpMarObjMarg1}-\eqref{eq:UpMarObjMarg4} by determining the $\gamma\ge 1$ most likely DA hypotheses with corresponding weights. For each DA, the joint posterior of the \ac{ue} state and landmarks is computed by the EKF approximation around the prior mean. These posteriors are marginalized and finally fused according to their weights. This leads to an efficient implementation, amenable for near real-time implementation.

\section{Posterior Linearization}
In the PMB(M) SLAM filter, we need to update the state of \ac{ue} and the map jointly for each DA. To linearize the nonlinear measurement function of the joint state of the \ac{ue} and landmarks is important. In this section, we will introduce the basics of linearization, argue that linearization should be done with respect to the posterior \ac{PDF} instead of at the prior mean, and propose a method to realize the posterior linearization. We drop the time index and the DA index for simplicity. Hence, we can denote the joint state of the \ac{ue} and landmarks given a certain DA as $\check{\boldsymbol{s}}$, the corresponding measurement function as $\check{\boldsymbol{h}}(\check{\boldsymbol{s}})$, and the associated measurement vector as $\check{\boldsymbol{z}}$. More details can be found in \cite[Section. IV.D]{ge2021computationally}. 

\subsection{Linearization Principle}\label{section:linearization}
To linearize the nonlinear measurement function $\check{\boldsymbol{h}}(\check{\boldsymbol{s}})$ is to approximate it by a linear function with a zero-mean Gaussian noise,
%To address the linearity of the $\check{\boldsymbol{h}}(\check{\boldsymbol{s}})$, we do the approximation
as
\begin{align}
    \check{\boldsymbol{h}}(\check{\boldsymbol{s}}) \approx \boldsymbol{H}\check{\boldsymbol{s}}+\boldsymbol{b}+\boldsymbol{e}, \label{linearization}
\end{align}
where $\boldsymbol{H} \in \mathbb{R}^{d_{\check{\boldsymbol{z}}}\times d_{\check{\boldsymbol{s}}}}$ denotes the linearized matrix, $\boldsymbol{b} \in \mathbb{R}^{d_{\check{\boldsymbol{z}}}\times 1}$ denotes the bias vector, and  $\boldsymbol{e}\in \mathbb{R}^{d_{\check{\boldsymbol{z}}}\times 1}$ denotes a zero-mean Gaussian distributed variable with covariance matrix $\boldsymbol{\Omega}$ that is independent of $\check{\boldsymbol{s}}$ and the measurement noise. 
%Therefore, to linearize the nonlinear measurement function is to approximate it by a linear function with a zero-mean Gaussian noise. 
In other words,  to linearize  $\check{\boldsymbol{h}}$ is to select suitable $(\boldsymbol{H}, \boldsymbol{b}, \boldsymbol{\Omega})$ for \eqref{linearization} by minimizing a problem-specific objective function (see Section \ref{section:Posterior_linearization}). Once the linearization in \eqref{linearization} is performed,  measurement $\check{\boldsymbol{z}}$ follows $\mathcal{N}(\check{\boldsymbol{z}};\boldsymbol{H}\check{\boldsymbol{s}}+\boldsymbol{b},\boldsymbol{\Omega}+\check{\boldsymbol{R}})$, where $\check{\boldsymbol{R}}$ is the overall measurement noise covariance. Moreover, if  the prior follows $\mathcal{N}(\check{\boldsymbol{s}};\boldsymbol{m}^{-},\boldsymbol{P}^{-})$, the posterior \ac{PDF} becomes $\mathcal{N}(\check{\boldsymbol{s}};\boldsymbol{m}^{+},\boldsymbol{P}^{+})$ with \cite{garcia2015posterior}
\begin{align}
 &\boldsymbol{m}^{+}=\boldsymbol{m}^{-}+\boldsymbol{K}(\check{\boldsymbol{z}}-\boldsymbol{H}\boldsymbol{m}^{-}-\boldsymbol{b}),\label{eq:update_mean} \\
&\boldsymbol{P}^{+} = \boldsymbol{P}^{-}- \boldsymbol{K} \boldsymbol{H}\boldsymbol{P}^{-},
\label{eq:update_covariance}
\end{align}
where $\boldsymbol{K}=\boldsymbol{P}^{-}\boldsymbol{H}^{\text{T}}(\boldsymbol{H}\boldsymbol{P}^{-}\boldsymbol{H}^{\text{T}}+\boldsymbol{\Omega}+\check{\boldsymbol{R}})^{-1}$ is the Kalman gain.

\subsection{Prior and Posterior Linearization}\label{section:Posterior_linearization}
\begin{figure}%[t]
\center
\input{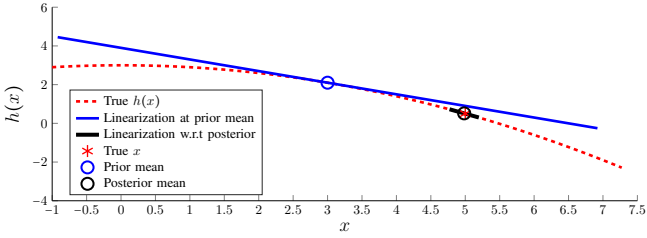}
\caption{An example of a nonlinear measurement function and its linearization at the prior mean and with respect to the posterior. The length of linearizations represents the 95\% confidence interval of the \acp{PDF}. The measurement function is set as $h(x)=-0.1x^{2}+3 +\eta$, where the variance of the measurement noise $\eta$ is 0.1. The prior follows a Gaussian distribution with mean as 3 and variance as 4, and we analyse the case where the measurement is 0.5.}
\label{Fig.linearization}
\end{figure}
\subsubsection{Prior Linearization with the EKF}
In the EKF, $\check{\boldsymbol{h}}(\check{\boldsymbol{s}})$ is approximated by the first-order Taylor series, and  \eqref{linearization} is linearized at the prior mean with $\boldsymbol{\Omega}=0$, and $\boldsymbol{H}$ representing the Jacobian of $\check{\boldsymbol{h}}(\cdot)$ with respect to $\check{\boldsymbol{s}}$, evaluated at the prior mean of $\check{\boldsymbol{s}}$. However, this approach does not make use of the measurement, thus its performance may deteriorate in some cases. Comparatively, it provides worse approximation than methods that use all available information, especially for nonlinear measurement functions with relatively low measurement noise \cite{garcia2015posterior}.

\subsubsection{Posterior Linearization}
To obtain a better approximation in \eqref{linearization}, the linearization should be done with respect to the posterior \ac{PDF} rather than at the prior mean \cite{garcia2015posterior}. The main reason is that the posterior \ac{PDF} is always narrower than the prior \ac{PDF}, especially if the measurement noise is low. Therefore, it is possible that linearization of the nonlinear measurement function at the prior mean makes the approximation lying outside the support of the  posterior \ac{PDF}. Fig.\,\ref{Fig.linearization} provides examples of linearization of a given measurement function at the prior mean and with respect to the posterior \ac{PDF}. It is obvious that the linearization with respect to the posterior \ac{PDF} provides better approximation and less uncertainty, as the linearization is more closer to the measurement function in the 95\% confidence interval, and the 95\% confidence interval is much shorter. The limitation is that the posterior is not yet available.

\subsubsection{Iterative Posterior Linearization Filter}\label{section: PLF}
A practical approach to implement posterior linearization is the \ac{IPLF}~\cite{garcia2015posterior}, %In Section \ref{section:Posterior_linearization}, we argued that the linearization should be done with respect to the posterior \ac{PDF}. 
which iteratively approximates $(\boldsymbol{H}, \boldsymbol{b}, \boldsymbol{\Omega})$ by solving the optimization problem \cite{garcia2015posterior}
\begin{align}
 \arg \underset{\boldsymbol{H}, \boldsymbol{b}} {\min} ~ &\mathbb{E} [  (\check{\boldsymbol{h}} - \boldsymbol{H}\check{\boldsymbol{s}}-\boldsymbol{b})^{\text{T}}(\check{\boldsymbol{h}} - \boldsymbol{H}\check{\boldsymbol{s}}-\boldsymbol{b})],\label{eq:optimizationProblem_Linearization} \\
\boldsymbol{\Omega} = &\mathbb{E} [  (\check{\boldsymbol{h}} - \boldsymbol{H}\check{\boldsymbol{s}}-\boldsymbol{b})(\check{\boldsymbol{h}} - \boldsymbol{H}\check{\boldsymbol{s}}-\boldsymbol{b})^{\text{T}}],
\label{eq:optimizationProblem_MSE}
\end{align}
where $\mathbb{E}[\cdot]$ represents the expectation with respect to the posterior \ac{PDF}. From \eqref{eq:optimizationProblem_Linearization},  we find the optimal $(\boldsymbol{H}, \boldsymbol{b})$ that can give the best linearization of $\check{\boldsymbol{h}}$ in the sense of minimizing its \ac{MSE}, and the corresponding \ac{MSE} matrix is then recovered as $\boldsymbol{\Omega}$ in \eqref{eq:optimizationProblem_MSE}.

\begin{figure}
\centering
    \includegraphics[width=0.8\linewidth]{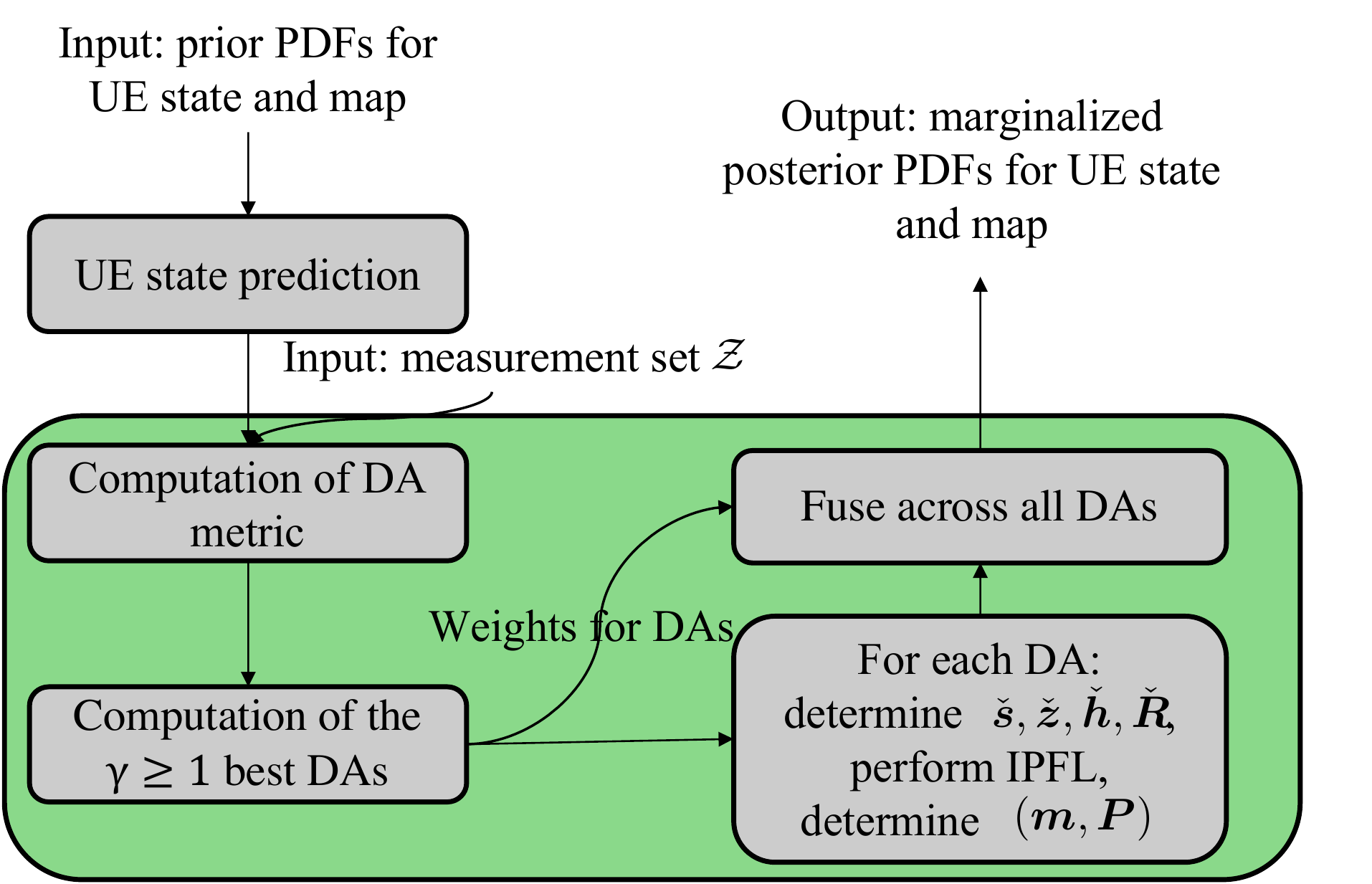}
\caption{The flowchart of the integration of the PMB SLAM filter from \cite{ge2021computationally} with the \ac{IPLF}. }
\label{fig:PMB_SLAM}
%\vspace{-4mm} 
\end{figure}

To solve this optimization problem, we can perform %\ac{SLR}  
\eqref{eq:optimizationProblem_Linearization} and \eqref{eq:optimizationProblem_MSE} with respect to iterative approximations of the posterior \ac{PDF}, starting from the prior \ac{PDF}. After each iteration, we obtain an improved approximation to the posterior \ac{PDF}, from which we can obtain an improved linearization. Given an approximation, the expectation in \eqref{eq:optimizationProblem_Linearization} can be computed using sigma point principle of the \ac{ckf} \cite{arasaratnam2009cubature}. %However, the posterior \ac{PDF} should be known to obtain the expectation in \eqref{eq:optimizationProblem_Linearization} and \eqref{eq:optimizationProblem_MSE} to approximate the posterior \ac{PDF}, which is problematic. Therefore, we can perform \ac{SLR} iteratively with respect to the best approximation to the posterior \ac{PDF}, denoted as \ac{IPLF}~\cite{garcia2015posterior}, and the intergral is approximated by using sigma point principle of the \ac{ckf} \cite{arasaratnam2009cubature}. 
The entire \ac{IPLF} procedure is summarized in Algorithm  \ref{alg:IPLF}.  
 The integration of the \ac{IPLF} is summarized in Fig.\,\ref{fig:PMB_SLAM}.

\begin{algorithm}[ht]
\caption{\ac{IPLF}} \label{alg:IPLF}
\begin{algorithmic}[1]
\Require \parbox[t]{\dimexpr\linewidth- \algorithmicindent * 1}{ Prior mean $\boldsymbol{m}$ and covariance $\boldsymbol{P}$;
\strut}
\Ensure Posterior mean $\boldsymbol{m}$ and covariance $\boldsymbol{P}$;
\Repeat
\State Factorize the covariance $\boldsymbol{P}$ by 
\begin{align}
     \boldsymbol{P} = \boldsymbol{G}  \boldsymbol{G}^{\text{T}} \label{factorize};
\end{align}
\For{$c\in\{1,\cdots,2d_{\check{\boldsymbol{s}}}\}$} %\algorithmiccomment{Each cubature point}
\State Compute cubature point
\begin{align}
     \check{\boldsymbol{s}}_{c} = \boldsymbol{G}\delta_{\check{\boldsymbol{s}},c} + \boldsymbol{m} \label{cubature_point};
\end{align}
\State Compute the propagated cubature point
\begin{align}
     \check{\boldsymbol{z}}_{c} = \check{\boldsymbol{h}}(\check{\boldsymbol{s}}_{c}) \label{cubature_point_pog};
\end{align}
\EndFor
\State Compute approximations of innovation $\tilde{\boldsymbol{z}}$, innovation covariance $\boldsymbol{S}_{\check{\boldsymbol{z}}\check{\boldsymbol{z}}}$ and  cross-covariance  $\boldsymbol{S}_{\check{\boldsymbol{s}}\check{\boldsymbol{z}}}$ by
\begin{align}
     &\tilde{\boldsymbol{z}} \approx \frac{1}{2d_{\check{\boldsymbol{s}}}} \sum_{c=1}^{2d_{\check{\boldsymbol{s}}}}\check{\boldsymbol{z}}_{c} \label{eq:sigma-z}, \\
     &\boldsymbol{S}_{\check{\boldsymbol{z}}\check{\boldsymbol{z}}} \approx \frac{1}{2d_{\check{\boldsymbol{s}}}} \sum_{c=1}^{2d_{\check{\boldsymbol{s}}}}(\check{\boldsymbol{z}}_{c}-\tilde{\boldsymbol{z}})(\check{\boldsymbol{z}}_{c}-\tilde{\boldsymbol{z}})^{\text{T}}\label{eq:sigma-Psi},\\
     &\boldsymbol{S}_{\check{\boldsymbol{s}}\check{\boldsymbol{z}}} \approx \frac{1}{2d_{\check{\boldsymbol{s}}}} \sum_{c=1}^{2d_{\check{\boldsymbol{s}}}}(\check{\boldsymbol{s}}_{c}-\boldsymbol{m})(\check{\boldsymbol{z}}_{c}-\tilde{\boldsymbol{z}})^{\text{T}}\label{eq:sigma-Phi};
\end{align}
\State Compute $(\boldsymbol{H}, \boldsymbol{b}, \boldsymbol{\Omega})$ by
\begin{align}
    &\boldsymbol{H} =  \boldsymbol{S}_{\check{\boldsymbol{s}}\check{\boldsymbol{z}}}^{\text{T}}\boldsymbol{P}^{-1}\label{eq:sigma-A},\\
& \boldsymbol{b} = \tilde{\boldsymbol{z}} -  \boldsymbol{H}\boldsymbol{m}\label{eq:sigma-b},\\
& \boldsymbol{\Omega} = \boldsymbol{S}_{\check{\boldsymbol{z}}\check{\boldsymbol{z}}} - \boldsymbol{H}\boldsymbol{P}\boldsymbol{H}^{\mathsf{T}}\label{eq:sigma-Omega};
\end{align}

\State Update $\boldsymbol{m}$ and $\boldsymbol{P}$ using \eqref{eq:update_mean} and \eqref{eq:update_covariance};
%\State $i=i+1;$
\Until {$\boldsymbol{m}$ and $\boldsymbol{P}$ converge \cite[eq.\,(30)]{garcia2015posterior}%Change in the posterior approximation is small enough
;}
\end{algorithmic}
Notation:  $\delta_{\check{\boldsymbol{s}},c}=\sqrt{d_{\check{\boldsymbol{s}}}}[\boldsymbol{I}_{d_{\check{\boldsymbol{s}}}\times d_{\check{\boldsymbol{s}}}},-\boldsymbol{I}_{d_{\check{\boldsymbol{s}}}\times d_{\check{\boldsymbol{s}}}}]_{1:d_{\check{\boldsymbol{s}}},c}$, with $\boldsymbol{I}_{d_{\check{\boldsymbol{s}}}\times d_{\check{\boldsymbol{s}}}}$ representing  a $d_{\check{\boldsymbol{s}}}\times d_{\check{\boldsymbol{s}}}$ identity matrix.
\end{algorithm}

\section{Results}
\subsection{Simulation Scenario}
Simulations are performed for a  5G application scenario at 28 GHz with a single known \ac{bs} and an unknown vehicle, which does a  counterclockwise  constant turn-rate movement around the \ac{bs}. The transmitter at the \ac{bs} side and the receiver at the vehicle side are both equipped with a uniform rectangular array (URA) with $8\times8$ antennas.  Every time step, the transmitter downlinks OFDM signals to the vehicle, with 16 symbols, 64 subcarriers, and 200~MHz bandwidths. Apart from the \ac{bs} and the vehicle,  there are  4  \acp{va}, and 4 \acp{sp} in the scenario. We implemented the \ac{IPLF} into the \ac{PMB} SLAM filter as in Fig. \ref{fig:PMB_SLAM}, which we denote as the IPL-\ac{PMB} SLAM filter. We compared the proposed IPL-\ac{PMB} SLAM filter with the EK-\ac{PMB} SLAM filter \cite{ge2021computationally}, which both consider the 10-best data associations every time step. We evaluated the mapping performance by the \ac{gospa} distance \cite{rahmathullah2017generalized} for both \acp{va} and \acp{sp}, and positioning performance  by the root mean squared error (RMSE) and standard deviation over time. We also measured the execution time of the two SLAM filters. More details and parameter settings can be found in  \cite{ge2021computationally}. The results were averaged over 100 Monte Carlo  simulations. All codes were written in MATLAB, and simulations were run on a MacBook Pro with a 2.6 GHz 6-Core Intel Core i7 processor and 16~Gb~memory.

\subsection{Results and Discussion}

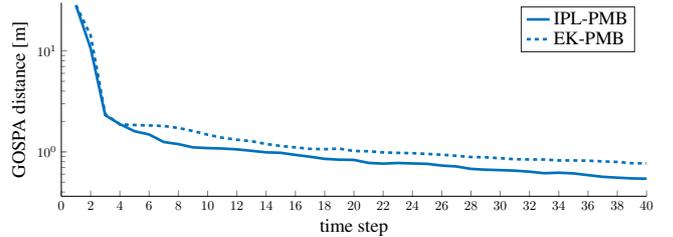
\begin{figure}
\center
\definecolor{mycolor1}{rgb}{0.00000,0.44700,0.74100}%
\definecolor{mycolor2}{rgb}{0.85000,0.32500,0.09800}%
\definecolor{mycolor3}{rgb}{0.00000,0.44700,0.74100}%
\definecolor{mycolor4}{rgb}{0.85000,0.32500,0.09800}%
\definecolor{mycolor5}{rgb}{0,0,0}%
%\definecolor{mycolor3}{rgb}{0.92900,0.69400,0.12500}%
%\definecolor{mycolor4}{rgb}{0.49400,0.18400,0.55600}%
%
\begin{tikzpicture}[scale=0.8\linewidth/14cm]

\begin{axis}[%
width=6.028in,
height=2.009in,
at={(1.011in,2.014in)},
scale only axis,
xmin=0,
xmax=40,
xlabel style={font=\color{white!15!black},font=\Large},
xlabel={time step},
ymin=0,
ymax=30,
ymode=log,
ylabel style={font=\color{white!15!black},font=\Large},
ylabel={GOSPA distance [m]},
axis background/.style={fill=white},
axis x line*=bottom,
axis y line*=left,
legend style={legend cell align=left, align=left, draw=white!15!black,font=\Large}
]

\addplot [color=mycolor1,  line width=2.0pt]
  table[row sep=crcr]{%
1	28.2842712474619\\
2	10.6233345707439\\
3	2.30684269740745\\
4	1.88263984798703\\
5	1.60054884481895\\
6	1.48634495936333\\
7	1.25206673923119\\
8	1.19252193025311\\
9	1.10772615254684\\
10	1.08929744281232\\
11	1.07878904264426\\
12	1.05830123104709\\
13	1.02398677084313\\
14	0.98942019944763\\
15	0.978188587933674\\
16	0.933195635208085\\
17	0.89476923698522\\
18	0.852286094195418\\
19	0.836953493769508\\
20	0.831818175155912\\
21	0.777608735544386\\
22	0.764761539580007\\
23	0.776408841182909\\
24	0.767805613569299\\
25	0.761129426378701\\
26	0.731603958829703\\
27	0.717048853839861\\
28	0.678033865446446\\
29	0.665876451536174\\
30	0.659291081997757\\
31	0.651610694542702\\
32	0.636332619691662\\
33	0.614329933139193\\
34	0.621661202740912\\
35	0.612053897895581\\
36	0.588276832958546\\
37	0.566118658486127\\
38	0.555309149133682\\
39	0.545857011752451\\
40	0.542785438684223\\
};
\addlegendentry{IPL-PMB}

\addplot [color=mycolor1,dashed, line width=2.0pt]
  table[row sep=crcr]{%
1	28.2842712474619\\
2	14.2815076877367\\
3	2.3608900273532\\
4	1.86958621565284\\
5	1.84516225700986\\
6	1.82471616888289\\
7	1.80336156394962\\
8	1.7231975543124\\
9	1.61050436045507\\
10	1.47896855142484\\
11	1.37991171497453\\
12	1.31889459290686\\
13	1.27495049797053\\
14	1.19680768138537\\
15	1.14339323656076\\
16	1.1069822560735\\
17	1.07043255541158\\
18	1.06273699026761\\
19	1.07699476815325\\
20	1.01864494481351\\
21	1.01429663711902\\
22	0.987275820613174\\
23	0.977165012472249\\
24	0.970575564873686\\
25	0.954105283182213\\
26	0.937263354530606\\
27	0.914164646178292\\
28	0.889351685504464\\
29	0.882919032535644\\
30	0.864814996967124\\
31	0.845338189598242\\
32	0.840632351631002\\
33	0.839442372262749\\
34	0.821967612893701\\
35	0.822412161423321\\
36	0.818034945505461\\
37	0.804953291279147\\
38	0.793965542821192\\
39	0.771694398539424\\
40	0.768904263367888\\
};
\addlegendentry{EK-PMB}

\end{axis}
\end{tikzpicture}%
\caption{Comparison of mapping performances for VAs between two SLAM filters.}
\label{Fig.mapping}
\end{figure}
\begin{figure}
\center
\definecolor{mycolor1}{rgb}{0.00000,0.44700,0.74100}%
\definecolor{mycolor2}{rgb}{0.85000,0.32500,0.09800}%
\definecolor{mycolor3}{rgb}{0.00000,0.44700,0.74100}%
\definecolor{mycolor4}{rgb}{0.85000,0.32500,0.09800}%
\definecolor{mycolor5}{rgb}{0,0,0}%
%\definecolor{mycolor3}{rgb}{0.92900,0.69400,0.12500}%
%\definecolor{mycolor4}{rgb}{0.49400,0.18400,0.55600}%
%
\begin{tikzpicture}[scale=0.8\linewidth/14cm]

\begin{axis}[%
width=6.028in,
height=2.009in,
at={(1.011in,2.014in)},
scale only axis,
xmin=0,
xmax=40,
xlabel style={font=\color{white!15!black},font=\Large},
xlabel={time step},
ymin=0,
ymax=30,
ymode=log,
ylabel style={font=\color{white!15!black},font=\Large},
ylabel={GOSPA distance [m]},
axis background/.style={fill=white},
axis x line*=bottom,
axis y line*=left,
legend pos=south west,
legend style={legend cell align=left, align=left, draw=white!15!black,font=\Large}
]

\addplot [color=mycolor1, line width=2.0pt]
  table[row sep=crcr]{%
1	28.2842712474619\\
2	24.4992372650315\\
3	24.4975289509128\\
4	24.4975289509128\\
5	24.4975289509128\\
6	24.4975289509128\\
7	24.4975289509128\\
8	20.0117074922972\\
9	20.0067563516343\\
10	20.0046232404385\\
11	20.004087414838\\
12	20.0044665663717\\
13	20.0045918063596\\
14	20.0045918063596\\
15	20.0045918063596\\
16	20.0045918063596\\
17	20.0045918063596\\
18	14.1562205429643\\
19	14.1491372994035\\
20	14.1502967154285\\
21	14.1515948935379\\
22	14.1509074833008\\
23	14.1510783246889\\
24	14.1510783246889\\
25	14.1510783246889\\
26	14.1510783246889\\
27	14.1510783246889\\
28	0.589460845839475\\
29	0.502421226108311\\
30  0.462781136477989\\
31	0.458099881580781\\
32	0.451084484312065\\
33	0.444890105676433\\
34	0.444890105676433\\
35	0.444890105676433\\
36	0.444890105676433\\
37	0.429010130372378\\
38	0.364871168828619\\
39	0.327169358562526\\
40	0.332004381004963\\
};
\addlegendentry{IPL-PMB}

\addplot [color=mycolor1, dashed, line width=2.0pt]
  table[row sep=crcr]{%
1	28.2842712474619\\
2	24.4976104126054\\
3	24.4971726800625\\
4	24.4971726800625\\
5	24.4971726800625\\
6	24.4971726800625\\
7	24.4971726800625\\
8	20.0089606935208\\
9	20.006432279241\\
10	20.0043857271464\\
11	20.0046877977079\\
12	20.0051328095331\\
13	20.0048052713288\\
14	20.0048052713288\\
15	20.0048052713288\\
16	20.0048052713288\\
17	20.0048052713288\\
18	14.1602334840715\\
19	14.1542186548884\\
20	14.1514978309904\\
21	14.1509867039574\\
22	14.1502552456764\\
23	14.1500692849031\\
24	14.1500692849031\\
25	14.1500692849031\\
26	14.1500692849031\\
27	14.1500692849031\\
28	0.637075649071575\\
29	0.514538379627771\\
30	0.492985276726793\\
31	0.505962405801442\\
32	0.512787304142804\\
33	0.500015157159195\\
34	0.500015157159195\\
35	0.500015157159195\\
36	0.500015157159195\\
37	0.499934267880708\\
38	0.400520670101598\\
39	0.373244048861521\\
40	0.388478532031204\\
};
\addlegendentry{EK-PMB}

\end{axis}
\end{tikzpicture}%
\caption{Comparison of mapping performances for SPs between two SLAM filters.}
\label{Fig.mapping_SP}
\end{figure}
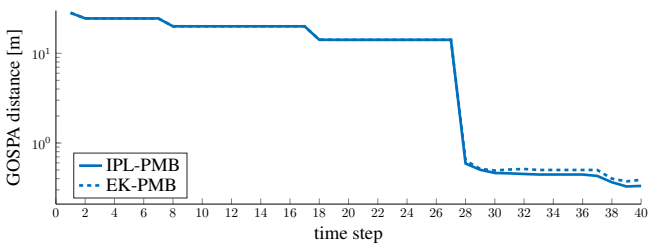

\begin{table}[]
    \centering
    \caption{Average standard deviations of the \ac{ue} state of the two SLAM filters.}
        \begin{tabular}{ |c|c|c|c|c| } 
         \hline
         Filter & $x$ [m] & $y$ [m] & heading [deg] & bias [m]\\ \hline 
         IPL-PMB  & $0.109$ & $0.109$ & $0.158$ & $0.086$
         \\ 
         EK-PMB   & $0.192$ & $0.201$ & $0.252$ & $0.151$
         \\ 
         \hline
        \end{tabular}
        
    \label{tab:covariance}
\end{table}
\begin{table}
    \centering
    \caption{Average computation time in milliseconds of the prediction and update steps of the two SLAM filters.}
        \begin{tabular}{ |c|c|c|c| } 
         \hline
         Filter & Prediction & Update & Total \\ \hline 
         IPL-PMB & $0.35$ & $28.1$ & $28.5$
         \\
         EK-PMB  & $0.34$ & $13.6$ & $13.9$
         \\ 
         \hline
        \end{tabular}
        
    \label{tab:cpu_time}
\end{table}

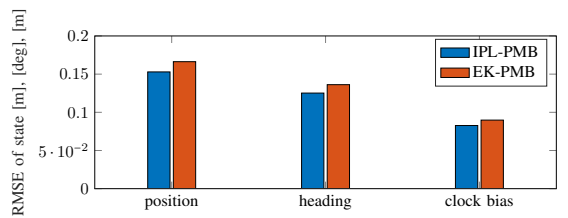
\begin{figure}
\center
% This file was created by matlab2tikz.
%
%The latest updates can be retrieved from
%  http://www.mathworks.com/matlabcentral/fileexchange/22022-matlab2tikz-matlab2tikz
%where you can also make suggestions and rate matlab2tikz.
%
\definecolor{mycolor1}{rgb}{0.00000,0.44700,0.74100}%
\definecolor{mycolor2}{rgb}{0.85000,0.32500,0.09800}%
\definecolor{mycolor3}{rgb}{0.92900,0.69400,0.12500}%
\definecolor{mycolor4}{rgb}{0.49400,0.18400,0.55600}
\definecolor{mycolor5}{rgb}{0,0,0}%
\begin{tikzpicture}[scale=0.99\linewidth/14cm]

\begin{axis}[%
width=3.842in,
height=1.281in,
at={(3.465in,2.378in)},
scale only axis,
bar shift auto,
xmin=0.5,
xmax=3.5,
xtick={1,2,3},
xticklabels={{position},{heading},{clock bias}},
ymin=0,
ymax=0.2,
ylabel style={font=\color{white!15!black}},
ylabel={RMSE of state [m], [deg], [m]},
axis background/.style={fill=white},
legend style={ anchor=north east, legend cell align=left, align=left, draw=white!15!black}
]

\addplot[ybar, bar width=0.145, fill=mycolor1, draw=black, area legend] table[row sep=crcr] {%
1	0.1529\\
2	0.1252\\
3	0.0827\\
};
\addplot[forget plot, color=white!15!black] table[row sep=crcr] {%
0.5	0\\
3.5	0\\
};
\addlegendentry{IPL-PMB}

\addplot[ybar, bar width=0.145, fill=mycolor2, draw=black, area legend] table[row sep=crcr] {%
1	0.1663\\
2	0.1362\\
3	0.0898\\
};
\addplot[forget plot, color=white!15!black] table[row sep=crcr] {%
0.5	0\\
3.5	0\\
};
\addlegendentry{EK-PMB}

\end{axis}
\end{tikzpicture}%
\caption{Comparison of \ac{ue} state estimation between two SLAM filters.}
\label{Fig.bar}
\end{figure}

Fig.~\ref{Fig.mapping} and Fig.~\ref{Fig.mapping_SP} shows the comparison of GOSPA results between the proposed IPL-\ac{PMB} SLAM filter with the EK-\ac{PMB} SLAM filter for \acp{va} and \acp{sp}, respectively. 
We observe that solid lines are below dashed lines in both figures, which shows the proposed IPL-\ac{PMB} SLAM filter has  better mapping performance. 
The reason is that the linearization is done at the prior mean in the EK implementation, which does not use the information provided by the measurement and provides worse approximation.  Unlike the EK implementation ignores the measurement in linearization, the proposed IPL-\ac{PMB} filter makes use of the measurement, and linearizes the measurement function with respect to the posterior \ac{PDF}, which provides more accurate and precise approximation to the measurement function, as we discussed in Section \ref{section:Posterior_linearization}. This is also the reason why the IPL-\ac{PMB} filter has slightly better accuracy in positioning performance, as its RMSEs are lower in Fig.~\ref{Fig.bar}. Although the accuracy is not improved significantly, the IPL implementation provides much more precise results,  as standard deviations of the \ac{ue} state decrease to half of what EK-\ac{PMB} SLAM filter provides approximately, shown in Table \ref{tab:covariance}.

Table \ref{tab:cpu_time} displays the execution time of the two SLAM filters. Since two algorithms have the exactly same prediction step, the prediction time is nearly identical. The proposed IPL-\ac{PMB} SLAM filter takes longer time than the EK-\ac{PMB} SLAM filter in the update step, with 28.1 ms and 13.6 ms per time step, respectively. The reason is that the linearization is done iteratively, and the sigma point principle is used to approximate innovation, innovation covariance, and cross-covariance between state and innovation in Algorithm \ref{alg:IPLF}, while the EKF directly linearizes the measurement function by using the first-order Taylor  series at the prior mean. This leads to $(2d_{\check{\boldsymbol{s}}} \times A)$-fold complexity in updating the joint state under each DA, where $A$ represents the number of iterations and is 5.3 on average over time and DAs in our implementation. Although the IPL-\ac{PMB} takes longer time, online and real-time operation of the filter could still be  guaranteed  with  more  accurate and precise  performance.

\section{Conclusions}
In this paper, we have provided the update details of the \ac{IPLF}, implemented the posterior linearization into the \ac{PMB}-based \ac{slam} filter in a 5G downlink scenario, which utilizes the measurement to linearize the measurement model with respect to the posterior \ac{PDF},  and proposed the  IPL-\ac{PMB} SLAM filter. Via simulation results, we demonstrate that the proposed IPL-\ac{PMB} SLAM filter is the same as the EK-\ac{PMB} SLAM filter that can map the environment and estimate the \ac{ue} simultaneously. Our results also indicate that the implementation of the posterior linearization helps the \ac{PMB} SLAM filter acquire more accurate and  precise estimates. Although additional computational cost is needed to obtain such performance gain, online and near real-time operation of the filter could still be  guaranteed.

\section*{Acknowledgment}
{This work was partially supported by the Wallenberg AI, Autonomous Systems and Software Program (WASP) funded by Knut and Alice Wallenberg Foundation, and the Vinnova 5GPOS project under grant 2019-03085, by the Swedish Research Council under grant 2018-03705.}
\balance

\bibliography{IEEEabrv,Bibliography}

% trigger a \newpage just before the given reference
% number - used to balance the columns on the last page
% adjust value as needed - may need to be readjusted if
% the document is modified later
% \IEEEtriggeratref{7}
% The "triggered" command can be changed if desired:
% \IEEEtriggercmd{\enlargethispage{-20cm}}

% references section

% can use a bibliography generated by BibTeX as a .bbl file
% BibTeX documentation can be easily obtained at:
% http://mirror.ctan.org/biblio/bibtex/contrib/doc/
% The IEEEtran BibTeX style support page is at:
% http://www.michaelshell.org/tex/ieeetran/bibtex/
%\bibliographystyle{IEEEtran}
% argument is your BibTeX string definitions and bibliography database(s)
%\bibliography{IEEEabrv,../bib/paper}
%
% <OR> manually copy in the resultant .bbl file
% set second argument of \begin to the number of references
% (used to reserve space for the reference number labels box)

% that's all folks
\end{document}